\documentclass[showpacs,floatfix,eqsecnum,nofootinbib]{revtex4}  
\usepackage{epsfig}

\begin{document}
\title{$B\to\rho$ form factors including higher twist
  contributions and reliability of pQCD approach}

\author{Namit Mahajan}
\email{nmahajan@mri.ernet.in}
\affiliation{{\em Harish-Chandra Research Institute,} \\
	 {\em Chhatnag Road, Jhunsi, Allahabad - 211019, India.}}

\def\be{\begin{equation}}
\def\ee{\end{equation}}
\def\bea{\begin{eqnarray}}
\def\eea{\end{eqnarray}}
\def\ba{\begin{array}}
\def\ea{\end{array}}
\def\prho{P_{\rho}}
\def\pb{P_B}
\def\kperp{k_{\perp}}
\def\lperp{l_{\perp}}
\def\btorho{B\to\rho}
\def\mb{m_B}
\def\mrho{m_{\rho}}
\def\fb{f_B}
\def\frho{f_{\rho}}
\def\fperp{f_{\perp}}
\def\deltal{\triangle(l)}
\def\phiparallel{\phi_{\parallel}(x)}
\def\hparallelt{h^{(t)}_{\parallel}(x)}
\def\hparallelsprime{h^{\prime(s)}_{\parallel}(x)}
\def\hparallels{h^{(s)}_{\parallel}(x)}
\def\phiperp{\phi_{\perp}(x)}
\def\gperpv{g^{(v)}_{\perp}(x)}
\def\gperpaprime{g^{\prime(a)}_{\perp}(x)}
\def\gperpa{g^{(a)}_{\perp}(x)}
\def\psib{\Psi_B(\xi)}
\def\psibbar{{\bar{\Psi}}_B(\xi)}
\def\psibplus{\Psi_B^+}
\def\psibminus{\Psi_B^-}
\def\phibplus{\phi_B^+}
\def\phibminus{\phi_B^-}
\def\sigmab{\Sigma_B}
\def\sigmarho{\Sigma_{\rho}}
\def\wb{\omega_B}
\def\wrho{\omega_{\rho}}
\def\bb{b_B}
\def\brho{b_{\rho}}
\def\dellperp{\partial^{l_{\perp}}}
\def\delkperp{\partial^{k_{\perp}}}
\def\ep{\epsilon^*}
\def\epslash{\not{\epsilon^*}}
\def\nplus{n_+}
\def\nminus{n_-}
\def\nplusslash{\not{n_+}}
\def\nminusslash{\not{n_-}}
\def\epperp{\epsilon_{\perp}^*}
\def\epperpslash{\not{\epsilon_{\perp}^*}}
\def\qsq{q^2}
\def\qhat{\hat{q}}
\def\bhat{\hat{b}}
\def\fac{\Bigg(1-\frac{\eta^2}{2(\eta^2-r^2)}\Bigg)}

\begin{abstract}
We discuss $\btorho$ form factors within the framework of
perturbative QCD, including the higher twist contributions 
and study the validity of such an approach in
calculating quantities such as form factors, which in principle and
quite generally are thought to be completely non-perturbative objects
and are expected to receive large contributions from the
non-perturbative regime in the calculations. It is shown that
including the Sudakov and threshold resummation effects, the general
expectations of the pQCD approach are clearly met and the form factors
do indeed receive most of the contribution from the perturbative
region. We do not make an attempt to precisely evaluate the form
factors but rather try to study the gross features and behaviour of
the same. We also find that use of single wave function for the B-meson
may actually underestimate various quantities. The results clearly
indicate the validity and reliability of pQCD calculations, at least in
this particular case.    
\end{abstract}
\maketitle

\begin{section}{Introduction}
\indent B-decays offer a very fertile soil for understanding,
analysing and testing the basic structure of particle interactions, widely
ranging from electroweak aspects, including CP violation, to the
mysterious world of QCD (for a quick review of various issues see
\cite{ali}).
 The study of B-decays and related observables
has greatly enhanced and affected our understanding of the underlying
principles that guide and govern these phenomena. The Standard Model
(SM) of particle physics, including the QCD corrections, seems to be a
highly successful candidate in explaining almost all the experimental
data, including a variety of issues concerning B-decays
themselves. However, as we enter the precision era in B-physics, there
is a compelling need to uncover more and more that goes as input while
evaluating and analysing these decays, and as can be expected, the
need to unearth the relative importance of QCD corrections is
overwhelming. 
In particular, the experience with $b\to s \gamma$ shows
that QCD can, in fact, alter the results by a large amount and
therefore it becomes imperative to include higher order QCD
corrections to get more and more sensible and accurate results.\\
\indent Semileptonic decays of the B-mesons are of particular interest
owing to their cleanliness and relatively simpler calculational
aspects. It also means that these decays provide us with the
opportunity to test QCD corrections more reliably and much more easily,
compared to pure hadronic decay modes. This has been widely recognized
and a lot of work has been done in this direction. Owing to the large
mass of B-mesons, it is expected that the heavy quark effective theory
(HQET) description is a good one and the leading term is given simply
by the quark level process (see \cite{neubert}).
The sub-leading terms are suppressed by the
B-meson mass and in most of the applications, are therefore
neglected. Another simplifying assumption that is employed is the idea
of {\it factorization} \cite{fac}. Quite simply, it means that in an energetic
process, like the decay of a B-meson, since the energy released is
large, the outgoing hadrons move out of the interaction region quite
quickly, thereby implying that there are no soft interactions between
the various hadronic subsystems. The full meson level amplitude for
any process is thus written as a convolution integral of the hard
scattering kernel (which can be reliably computed using standard
perturbative formalism) and the non-perturbative meson wave functions
that are universal and are obtained from lattice studies, fits to the
experimental data or sum rules of one kind or the other.
This assumption forms the backbone of
most of the calculations involving B-mesons. The issue of the
extent of reliability of such an approximation, which seems to work
quite well for most of the cases, is still an open question. It is
however found and argued that not in all cases this is a good
approximation and one must look for the sub-leading terms to this
approximation scheme. For example it suffers from the problem of scale
dependence \cite{li1} and in case of $B\to J/\Psi$ it seems difficult
to match the branching ratios using the approximation \cite{li2} 
 Furthermore, it is now accepted that though the
leading term in the HQET gives results that are quite close to the
observed numbers, the reality that the mass of a B-meson is not
infinite but close to $5$ GeV, implies that the sub-leading terms
can not be naively neglected anymore. One can try and estimate these
sub-leading terms in HQET itself or employ some other methods.\\
\indent It is worthwhile to try and explore the idea of estimating the
above mentioned sub-leading terms suppressed by the B-meson mass. To do such
a calculation, one can follow either of the following two
approaches. Both the approaches treat the hard scattering kernels
perturbatively and the mesonic wave functions are the non-perturbative
ingredients. The main difference lies in the treatment of quantities
like from factors.  \\
\begin{enumerate}
\item[(a)] {\it QCD factorization}: The form factors are also thought to be
  purely non-perturbative objects, not calculable perturbatively \cite{bbns}.        
\item[(b)] {\it pQCD}: It is believed that the form factors can be reliably
  and satisfactorily computed in perturbation theory. This approach
  was developed by Brodsky and Lepage and others \cite{pqcd}. The
  factorization theorem for relevant for exclusive B-meson has been
  proved \cite{brodsky}.
\end{enumerate}
There is no clear consensus on either of these two methods and both
seem to have some advantageous features while both suffer from lack of
very sound theoretical footing on certain issues. Both the approaches
have invited intense activity in the recent past. In some cases, the
results of the two approaches tend to be consistent while in some
other cases, they seem to be distance apart.\\
\indent In this note, we investigate $\btorho$ form factors in the
second of the approaches listed above, namely the pQCD method. We have
chosen semileptonic process for the merit of cleanliness and
relatively simplified calculations and the fact that with just two
hadrons in the process, we do not have to bother about the subtleties
and complications involved in a purely hadronic process and the
reliability can be checked more clearly. The same process has been
considered by \cite{hwang} and \cite{sanda}. 
We extend their study and elaborate on
the differences in the next sections.
The aim of this study is not to precisely pin down the value of the form
factors from the calculations but to explore and study the behaviour
of the same with respect to the various parameters that enter the
calculation and check the reliability of such a calculation. In
particular, we keep in mind the objections and criticisms that are
generally raised regarding the validity of such a scheme and try to
see for ourselves, whether, following a more consistent treatment
compared to whatever exists in literature till now, we get a clue to
some of the unresolved and mysterious issues that we are forced to
live with in such calculations. We would like to emphasize again that
the goal is not a very accurate numerical study of the form factors
but to study the general behaviour for some suitable choices of
various parameters like the shape variables appearing in the meson
wave functions etc. Furthermore, we would like to remind ourselves
that in such a situation, we may finally end up over- or
under-estimating some of the quantities but we hope to clarify certain
issues in the end. In this spirit, this study aims at extending the
ongoing debate between the two approaches in order to get a clearer
picture of what exactly is happening and can we understand and explain
the same. The article is organized as follows: in the next section we
briefly review the chief ingredients of the pQCD approach. We
summarize some of the main
objections/criticisms that this method faces. Next we discuss the form
factor calculation including the higher twist contributions to the
meson wave functions. We then qualitatively study the behaviour of the
various form factors without attempting to determine the values very
precisely. The last section discusses and summarizes our results and
conclusions. 
\end{section} 
\begin{section}{pQCD approach - A quick look}
\indent Exclusive processes enjoy the simplicity and elegance they derive
from {\it factorization theorems} invoked in some form or the
other (for a general review of factorization theorem and its related
issues see \cite{collins}). 
It has been demonstrated that for an exclusive process involving
a large scale $Q$, the
amplitude can be neatly written as
\be
{\mathcal{A}} \sim C(t)\otimes
H(t)\otimes\left(\prod_i\Phi_i(x_i)\right)\otimes exp[-S]
\ee
where $C(t)$ denotes the Wilson coefficients relevant to the problem,
$H(t)$ is the hard scattering kernel that is perturbatively evaluated,
$\Phi_i(x)$ describes the distribution of the partons in the $i$-th hadron
(here $x$ is the momentum fraction carried by the parton) and we
have collectively put all the resummed quantities in the factor $S$
for simplicity. The factor $S$ therefore contains the Sudakov
logarithms and the relevant evolution factor (these are discussed
below). The factorization theorem implies that in a hard exclusive
process, the non-perturbative dynamics can be separated from the
perturbative pieces and the final result is simply the convolution of
these. We now concentrate on a specific process involving B-meson
decay, namely $B\to\rho\ell\nu$, as our prototype hard exclusive
process. The large mass of the B-meson acts as the natural large scale
in the problem. Here and in the following we do not differentiate
between a B or a ${\bar{B}}$-meson.
We therefore have the following scales in the problem:
mass of the b-quark (we do not differentiate between the b-quark mass and
the B-meson mass) $m_b\sim 5$ GeV, the W-boson mass $M_W\sim 100$ GeV, 
the renormalization
scale $\mu$ (which is of the order of $m_b$) and a scale of the order of
$\Lambda_{QCD} \equiv \Lambda\sim250$ GeV. The Wilson coefficients
appearing in the effective Hamiltonian contain the information of the
high scales ($>m_b$) and have the resummed logarithms of the form
$\ln(M_W^2/\mu^2)$. The scale $\Lambda$ characterizes the
non-perturbative scale and crucially enters the meson wave
functions. The precise shape and behaviour of the meson wave function
are crucial inputs for any calculation to make physical sense. For
example, if the wave function does not vanish at the end points ($x\to
0,1$), then it can be shown that the amplitude is infra-red
divergent \cite{isgur}. 
Also, it has been pointed out that if the small transverse
momentum of the parton, generically denoted as 
$\kperp\sim{\mathcal{O}}(\Lambda)$ is ignored,
the dominant contribution arises from the end point regions. The way
out is to retain the $\kperp$ components, which in turn regulate
the infra-red divergence that has just been mentioned \cite{sterman}.
 If the variable
conjugate to $\kperp$ is denoted by $b$, then apart from the large
logarithms that are reorganised using standard renormalization group
techniques into Wilson coefficients, we have another potential source
of large logarithms - $\ln(\mu b)$. The inclusion of the transverse
momentum for the partons in turn results in double logarithms of the
form $\ln^2(Pb)$, where $P$ is the momentum of the meson (generally the
larger of the light-cone momentum components). These large logarithms are
resummed and lead to the so called Sudakov form factor
$exp[-S(p,b)]$. This exponentially damping factor suppresses the
long-distance contributions from the large $b$ regions and the factor
$S(P,b)$ vanishes as $b\to 1/\Lambda$. Since the infra-red divergences
in a theory are manifestations of the non-perturbative dynamics, such
effects are absorbed in the wave functions. These wave functions are
universal in character and do not depend on the specific
process. After, reorganising all the large logarithms and absorbing
the infra-red divergences in the meson wave functions, we are left
with finite quantities that are believed to be perturbatively
calculable. The main philosophy behind the pQCD method is that all the
factors except the hadronic wave functions can be perturbatively
calculated such that the dominant contribution to various quantities,
like form factors, comes from the hard gluon exchanges. It is
important that this is true because otherwise the dominant
contribution would arise from the non-perturbative regime - a region
where the perturbative calculations do not make sense and there is
absolutely no way of determining them in this way.  \\
\indent The $\btorho$ form factors are parameterized as follows \cite{ball1}:
\bea
\langle\rho (\prho,\lambda)\vert(V-A)_{\mu}\vert B(\pb)\rangle &=&
-i(\mb+\mrho)A_1(\qsq){\ep_{\mu}}^{(\lambda)} ~+~
\frac{iA_2(\qsq)}{(\mb+\mrho)}(\ep\cdot\pb)(\pb+\prho)_{\mu} \\ \nonumber
&+& \frac{iA_3(\qsq)}{(\mb+\mrho)}(\ep\cdot\pb)(\pb-\prho)_{\mu} ~+~
\frac{2V(\qsq)}{(\mb+\mrho)}\epsilon_{\mu\nu\alpha\beta}\pb^{\nu}
\prho^{\alpha}{\epperp}^{\beta}
\eea
where $\qsq = (\pb-\prho)^2$ is the momentum transferred.
For the process under consideration, we have the following 
two diagrams contributing at the one gluon exchange level in pQCD:
\vskip 1.0cm
\begin{figure}[ht]
\vspace*{-1cm}
\centerline{
\epsfxsize=7.0cm\epsfysize=2.5cm
                      \epsfbox{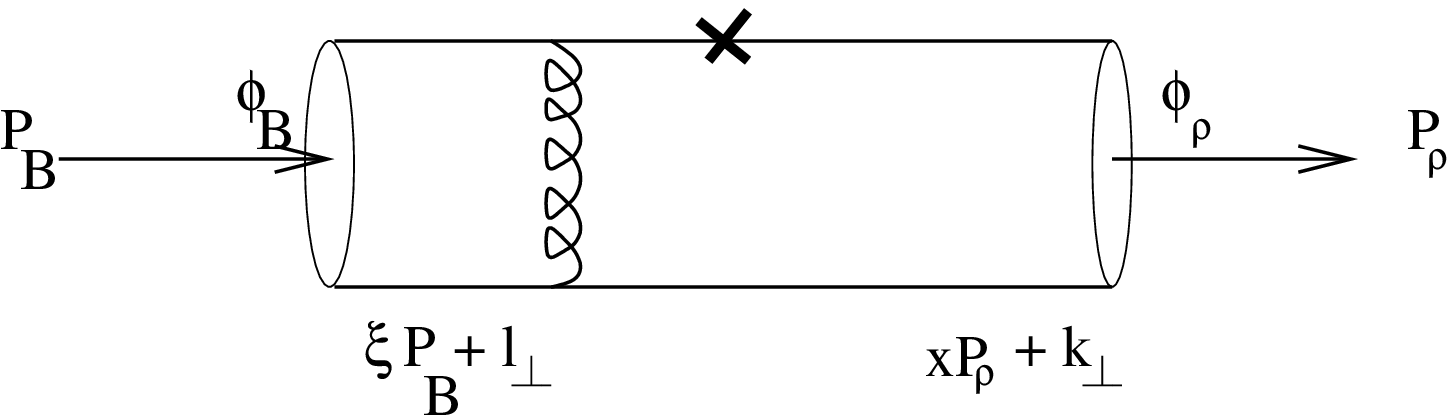}
\hskip 1.0cm 
\epsfxsize=7.0cm\epsfysize=2.5cm
                      \epsfbox{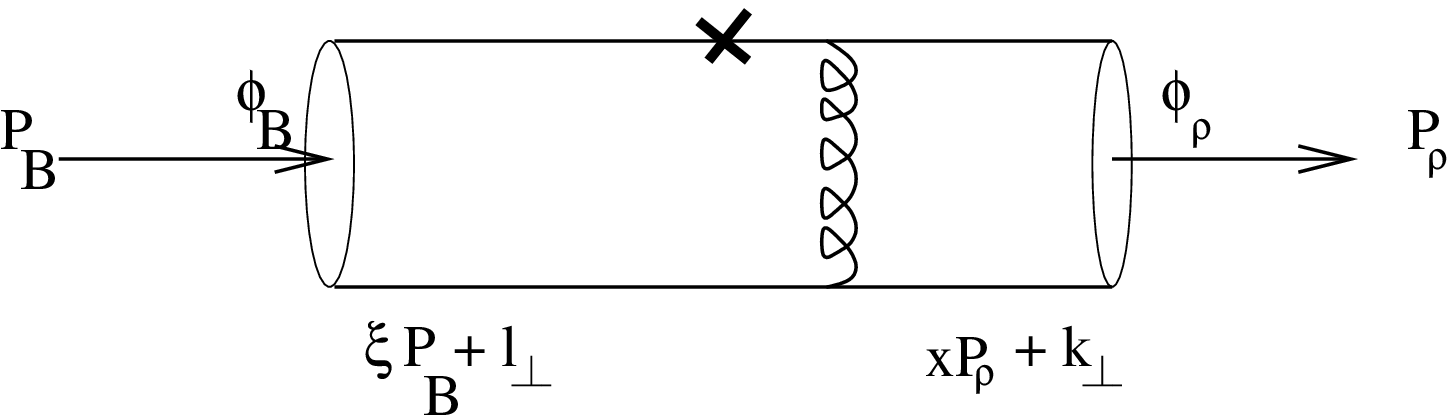}}
\caption{One gluon exchange diagrams contributing to $\btorho$ form
  factors. The cross denotes the weak interaction vertex.}
\end{figure}
where $\phi_{B,\rho}$ represents the corresponding meson wave
function, $\xi$ is the momentum fraction carried by the lighter
(anti)quark and $\lperp$ is the transverse momentum while $x$ and
$\kperp$ denote the analogous quantities for the $\rho$-meson. As
mentioned before, we do not distinguish between b-quark mass and
B-meson mass and set them equal in the calculations. The light quarks -
$u,d,s$ are taken to be massless. Our convention for the light-cone
variables is as follows. For any four momentum
$a^{\mu}=(a^+,a^-,\bf{a_{\perp}})$  we define
$a^{\pm} = a^0\pm a^3$ and ${\bf{a_{\perp}}}=(a^1,a^2)$. The scalar
product is defined as: $a\cdot
b=\frac{a^+b^-+a^-b^+}{2}-\bf{{a_{\perp}}}\cdot\bf{{b_{\perp}}}$. With
these conventions, we have for various momenta:
\be
\pb = \mb(1,1,{\bf{0_{\perp}}}) \hskip 1cm \prho =
\frac{\mb}{\eta}(r^2,\eta^2,{\bf{0_{\perp}}}) 
\ee
where $r=\mrho/\mb$ and $\eta=1-\qsq/\mb^2$. It is convenient to
define two light-like momenta $\nplus=(\sqrt{2},0,\bf{0_{\perp}})$
and $\nminus=(0,\sqrt{2},\bf{0_{\perp}})$ such that $\nplus\cdot\nminus=1$.   
The transverse momenta $\kperp$ and $\lperp$ as introduced above
should be thought of as four-vectors with only the transverse
components non vanishing. Schematically, therefore, the various form
factors (generically called $F$), after using the factorization
theorem, can be written as
\be
F = \int_0^1dx\int_0^1d\xi~\phi_B(\xi,\qsq)H(x,\xi,\qsq)\phi_{\rho}(x,\qsq)
\ee
where as before $\phi$'s denote the meson wave functions and
$H(x,\xi,\qsq)$ represents the hard kerne evaluated
perturbatively. The basic theme behind such a factorization is that
the long-distance or the soft interactions occur before and after the
hard decay process and therefore the two effects simply decouple. The
hard contribution takes place at short-distance scales and therefore we
only need to specify the distribution of quarks (partons) inside the
mesons apart from explicitly computing the perturbative piece. The
information regarding the parton distribution, along with all the
infra-red dynamics, is contained in the wave functions. In the
asymptotic region, the factorization theorem has been shown to be
valid. However, it has been pointed out \cite{isgur} that in the case
of pion's electromagnetic form factor at lower energy scales (few GeV's), 
large contributions come from
the end point regions ($x,\xi\to 0$). The perturbative analysis is not
valid in such a region and this casts serious doubts about such a
method. This problem, even when the total convolution integral does
not contain any divergences, is called the {\it endpoint problem}.
The way out is to introduce Sudakov factors that
regulate this undesirable endpoint behaviour and thus make the pQCD
calculations sensible at these lower energy scales \cite{botts}. The Sudakov
factors typically suppress the long-distance effects at large
transverse distances (denoted by variable $b$ earlier) ie. small
transverse momenta.
Therefore, the
endpoint behaviour is a crucial aspect to get any meaningful results.\\
\indent A further issue of concern is that as the momentum fraction
carried by the spectator tends to zero, the form factors become
divergent due to divergent behaviour of the meson wave functions
related to the light-cone distribution amplitudes (LCDAs). 
There is a need to resum the
contributions of the form $\ln^2x$. This is achieved by threshold
resummation into a jet function $S_t(x)$ such that $S_t(x) \to 0$ as
$x\to 0,1$ \cite{li3}. 
Therefore, the threshold resummation modifies the endpoint
behaviour of the distribution amplitudes (DAs).\\
\indent After including all these factors in the DAs,
one expects that the pQCD calculations are reliable and
can be carried out without much trouble. However, the method still
faces some more objections/criticisms which we now briefly discuss. We
summarize the main issues as discussed in \cite{sachrajda}\\
\begin{enumerate}
\item[$\bullet$] Are Sudakov suppression factors strong enough to
  regulate the large transverse separation contributions? In
  particular, it is quite possible that for intermediate values of the
  variables, away from the endpoints, there can be a slight
  enhancement. How effective and efficient is this effect?
\item[$\bullet$] How small are the contributions from the
  non-perturbative regime? 
\item[$\bullet$] The biggest uncertainties presumably arise from the
  precise lack of knowledge about the meson wave functions. In
  particular, the use of a single wave function for the B-meson has
  been questioned.
\end{enumerate} 
All these and related issues cast a serious cloud of doubt on the
applicability and reliability of pQCD calculations. We try to address
and clarify some of these issues below and also elaborate on various issues.
\end{section} 
\begin{section}{Meson distribution amplitudes}
\indent The DAs refer to the distribution of partons inside a hadron
(here mesons). A DA thus, carries the information regarding the
momentum fraction carried by partons inside the meson in a specific
Fock state. The DA is related to the the Bethe-Salpeter wave function
by the following relation
\be
\phi(x) \sim \int^{\vert\kperp\vert<\mu}d^2\kperp\phi_{BS}(x,\kperp)
\ee 
where $\mu$ is the ultra-violet cut-off (for a general discussion see
\cite{lepage}. In the discussions below, we
do not remain very careful and do not distinguish between the
wave functions and DAs and freely interchange one for the other
in the discussions.\\ 
\indent We closely follow the approach and results for the $\rho$-meson DAs as
outlined in \cite{ball2}. Accordingly, the the LCDAs are defined
as meson-to-vacuum transition matrix elements of the non-local gauge
invariant operators in the light-cone picture. For the light vector
mesons, the DAs are split into chiral even and chiral odd
contributions. In terms of
$n_{\pm}$, the vector meson momentum is written as $\prho = E\nminus +
\mrho^2{\nplus}_{\mu}/(4E)$, such that the transverse plane is defined
with respect to the vectors $n_{\pm}$. Fourier transforming the DAs,
we obtain the momentum space representation of the $\rho$-meson
light-cone projection \cite{beneke}
(keeping terms to twist-3 of the two particle
quark-antiquark distribution only and not writing the path ordered
integral and also remembering a relative $-i$ factor between the
definitions here and those in \cite{ball2})
\be
\langle\rho
(\prho,\lambda)\vert\bar{d}_{\alpha}(z)u_{\delta}(0)\vert0\rangle
\equiv M^{\rho}_{\delta}{\alpha} = M^{\rho}_{\delta\alpha\parallel}
+ M^{\rho}_{\delta\alpha\perp}
\ee

where
\bea
M^{\rho}_{\delta\alpha\parallel} &=&
\Bigg(-\frac{i\frho}{4}\frac{\mrho(\ep\cdot\nplus)}{2E}
E\nminusslash\phiparallel ~-~
\frac{i\fperp}{4}\frac{\mrho(\ep\cdot\nplus)}{2E}
\Bigg[-\frac{i}{2}\sigma_{\mu\nu}\nminus^{\mu}\nplus^{\nu}\hparallelt
  \\ \nonumber
&& ~-~ iE\int_0^xdv[\phi_{\perp}(v) -
    h^{(t)}_{\parallel}(v)]\sigma_{\mu\nu}\nminus^{\mu}\delkperp_{\nu}
  ~+~ \frac{\hparallelsprime}{2}\Bigg]\Bigg)_{\delta\alpha} 
\eea

\bea
M^{\rho}_{\delta\alpha\perp} &=&
\Bigg(-\frac{i\fperp}{4}E\epperpslash\nminusslash\phiperp ~-~
\frac{i\frho\mrho}{4}\Bigg[\epperpslash\gperpv ~-~
  E\int_0^xdv[\phi_{\parallel}(v) -
    g^{(v)}_{\perp}(v)]\nminusslash{\epperp}^{\mu}\delkperp_{\mu} \\
  \nonumber
&& ~+~
   i\epsilon_{\mu\nu\rho\sigma}{\epperp}^{\nu}\nminus^{\rho}
\gamma^{\mu}\gamma_5\Bigg(\nplus^{\sigma}\frac{\gperpaprime}{8} - 
E\frac{\gperpa}{4}{\delkperp}^{\sigma}\Bigg)\Bigg]\Bigg)_{\delta\alpha} 
\eea
By $\delkperp$ we mean $\frac{\partial}{\partial\kperp}$ and we have
\be
{\epperp}^{\mu} \equiv {\ep}^{\mu} -
\frac{\ep\cdot\nplus}{2}\nminus^{\mu} -
\frac{\ep\cdot\nminus}{2}\nplus^{\mu}
\ee
Authors in \cite{hwang} and \cite{sanda} did not include 
the $\delkperp$ terms in their
calculations. We retain these terms as well and finally show that they
contribute significantly to some of the form factors. All the
individual distribution amplitudes are normalized to unity ie.
$\int_0^1dx~\phiparallel = 1$ etc. For the sake of completeness, we list
the individual distribution amplitudes \cite{ball2}:
\[
\phiparallel = 6x(1-x)[1 + 0.27(5[2x-1]^2-1)]
\]
\[
\hparallelt = 3(2x-1)^2 + 0.3(2x-1)^2[5(2x-1)^2 - 3] + 0.21[3 -
  30(2x-1)^2 +35(2x-1)^4]
\]
\[
\hparallelsprime \equiv \frac{\partial\hparallels}{\partial x} 
= 6(2x-1)[1 + 0.76(10x^2 - 10x + 1)]
\]
\be
\phiperp = 6x(1-x)[1 + 0.3(5(2x-1)^2-1)]
\ee
\[
\gperpv = \frac{3}{4}[1 + (2x-1)^2] + 0.24[3(2x-1)^2 - 1] + 0.12[3 -
  30(2x-1)^2 + 35(2x-1)^4]
\]
\[
\gperpa = 6x(1-x)[1 + 0.23(5[2x-1]^2-1)]
\]
\[
\gperpaprime \equiv \frac{\partial\gperpa}{\partial x} 
\]
Fourier transforming the wave functions with respect to the transverse
momentum leads us to the wave functions expressed in the transverse
separation $b$-space. The variable $b$ defines the transverse
separation between the quark and the anti-quark inside the meson. In
principle, the various DAs listed above can have different transverse
momentum dependence. But for simplicity, we assume same dependence for
all of them. This dependence is not known from first principles and
therefore, it is generally assumed that the full wave function,
including the transverse momentum dependence, is of the form
\be
\Psi(x,\kperp) = \phi(x)\Sigma(\kperp)
\ee 
such that apart from the normalization condition for $\phi$ listed
above, we have for the transverse part, 
$\int d^2{\bf {\kperp}}\Sigma(\kperp)=1$. The functional form of
$\Sigma(\kperp)$ is assumed to be a simple Gaussian distribution. In
the $b$-space, we therefore have
\be
\sigmarho(\brho) = \exp\Bigg(-\frac{\wrho^2\brho^2}{2}\Bigg)
\ee 
\indent For the heavy B-meson, following \cite{beneke} and \cite{wei}, 
we have the
following momentum space projection operator
\bea
M^B_{\delta\alpha} &=& -\frac{i\fb}{4}\Bigg((\not{\pb}+\mb)\Bigg[\psib
  + \frac{\nplusslash-\nminusslash}{2}\psibbar - \frac{1}{2}\deltal\gamma^{\mu}\dellperp_{\mu}\Bigg]\gamma_5\Bigg)_{\delta\alpha}
\eea
where $\psib$ and $\psibbar$ are the B-meson wave functions
corresponding to the DAs $\phi_B$ and $\bar{\phi}_B$ defined as
\be
\phi_B = \frac{\phi_B^++\phi_B^-}{2} \hskip 1.5cm
\bar{\phi}_B = \frac{\phi_B^+-\phi_B^-}{2}
\ee
The term with coefficient $\deltal$ works out to be proportional to
$l^+/\mb$ and is therefore generally dropped. We retain this as well
in the analysis and comment on its contribution later. Also, it has
been argued that the $\psibbar$ term is sub-leading compared to the
first term in the projection operator and is also dropped. This is one
of the major issues of debate concerning the pQCD calculations. In the
present study, we retain this sub-leading term also. Another reason of
retaining this term is the confusion whether the pseudoscalar and
axial matrix elements can both be described by the same DA in the
heavy quark limit. It has been pointed out that the two DAs considered
by the authors of \cite{li4} (and of subsequent works
 based on their proposal), 
do not satisfy the equations of motion and further raise
more doubts. In this study, we follow \cite{sachrajda} and use, as a
model, the following form for the two DAs which satisfy the equations
of motion
\be
\phibplus(\xi) = \sqrt{\frac{2}{\pi}}\frac{\xi^2\mb^2}{\wb^3}
\exp\Bigg[-\frac{\xi^2\mb^2}{2\wb^2}\Bigg] 
\ee
and
\be
\phibminus(\xi) = \sqrt{\frac{2}{\pi}}\frac{1}{\wb}
\exp\Bigg[-\frac{\xi^2\mb^2}{2\wb^2}\Bigg] 
\ee  
The quantity $\deltal$ introduced above is related to the DAs as
follows
\be
\deltal = \int_0^{l_+}dl(\phibminus - \phibplus)
\ee
\indent We next include the transverse momentum dependence for the 
B-meson through the following
\be
\sigmab(\bb) = \exp\Bigg[-\frac{\wb^2\bb^2}{4}\Bigg]
\ee
The question whether the above DAs are actually realistic and whether
they correctly reproduce all the features is an intriguing one and has
to be dealt separately. We therefore proceed with the assumption that
they do or at least capture the important features.
\end{section}
\begin{section}{Sudakov form factors and threshold resummation}
\indent The resummation of large transverse separation regions leads
to Sudakov form factors while the resummation over small fractional
momenta leads to threshold resummation.\\
\indent As mentioned earlier, the Sudakov factor
suppresses the double logarithms of the form $\ln^2Pb$ arising due to
the overlap of soft and collinear divergences. The parameter $b$ regulates
these kind of infra-red contributions. In terms of the variables
\be
\qhat \equiv \ln\Bigg[\frac{xQ}{\sqrt{2}\Lambda}\Bigg] \hskip 1.5cm 
\bhat \equiv \ln\Bigg[\frac{1}{b\Lambda}\Bigg]
\ee
the exponent of the Sudakov form factor, to NLO, reads \cite{li5}
\bea
s(x,b,Q) &=&
\frac{A^{(1)}}{2\beta_1}\qhat\ln\Bigg(\frac{\qhat}{\bhat}\Bigg) ~-~
\frac{A^{(1)}}{2\beta_1}(\qhat-\bhat) +
\frac{A^{(2)}}{4\beta^2_1}\Bigg(\frac{\qhat}{\bhat}-1\Bigg)
- \Big[\frac{A^{(2)}}{4\beta^2_1} ~-~
  \frac{A^{(1)}}{4\beta_1}\ln\Bigg(\frac{e^{2\gamma_E-1}}{2}\Bigg)\Bigg]\ln\Bigg(\frac{\qhat}{\bhat}\Bigg) \\ \nonumber
&+&
\frac{A^{(1)}\beta_2}{4\beta^3_1}\qhat\Bigg[\frac{\ln(2\qhat)+1}{\qhat}
  - \frac{\ln(2\bhat)+1}{\bhat}\Bigg] ~+~
\frac{A^{(1)}\beta_2}{8\beta^3_1}[\ln^2(2\qhat)-\ln^2(2\bhat)] \\
\nonumber
&+&
\frac{A^{(1)}\beta_2}{8\beta^3_1}\ln\Bigg(\frac{e^{2\gamma_E-1}}{2}\Bigg)
\Bigg[\frac{\ln(2\qhat)+1}{\qhat} -
  \frac{\ln(2\bhat)+1}{\bhat}\Bigg]\\ \nonumber
&-& \frac{A^{(2)}\beta_2}{16\beta^4_1}\Bigg[\frac{2\ln(2\qhat)+3}{\qhat} -
  \frac{2\ln(2\bhat)+3}{\bhat}\Bigg]
~-~
\frac{A^{(2)}\beta_2}{16\beta^4_1}\frac{\qhat-\bhat}{\bhat^2}[2\ln(2\bhat)+1]
\\ \nonumber
&+&
\frac{A^{(2)}\beta^2_2}{432\beta^6_1}\frac{\qhat-\bhat}{\bhat^3}
[9\ln^2(2\bhat)+6\ln(2\bhat)+2] \\ \nonumber
&+&
\frac{A^{(2)}\beta^2_2}{1728\beta^6_1}\Bigg[\frac{18\ln^2(2\qhat)+
30\ln(2\qhat)+19}{\qhat^2} -
  \frac{18\ln^2(2\bhat)+30\ln(2\bhat)+19}{\bhat^2}\Bigg]
\eea
where $\gamma_E$ is the Euler's constant and the various coefficients
are given as follows:
\be
\beta_1 = \frac{33 - 2n_f}{12} \hskip 1.5cm \beta_2 = \frac{153 -
  19n_f}{24}
\ee
\be
A^{(1)} = \frac{4}{3} \hskip 1.5cm A^{(2)} = \frac{67}{9} -
\frac{\pi^2}{3} - \frac{10}{27}n_f +
\frac{8}{3}\beta_1\ln\Bigg(\frac{e^{\gamma_E}}{2}\Bigg) 
\ee
with $n_f$ being the number of flavours. The strong coupling constant
to NLO is given by
\be
\frac{\alpha_s(\mu)}{\pi} = \frac{1}{\beta_1\ln(\mu^2/\Lambda^2)} -
\frac{\beta_2}{\beta^3_1}\frac{\ln\ln(\mu^2/\Lambda^2)}{\ln^2(\mu^2/\Lambda^2)}
\ee
The Sudakov factor therefore falls for large $b$ regions and vanishes
as $b>1/\Lambda$. Also, the wave functions (or equivalently DAs)
defined above are valid only for scales $1/b$. To have the correct
results at an arbitrary $\mu$ a RG-evolution is required which
generates an evolution factor in the exponential. We club this factor
along with the Sudakov exponent and define the Sudakov as $exp(-S)$
with 
\be
S(x,b,Q,\mu) = s(x,b,Q) + s(1-x,b,Q) -
\frac{1}{\beta_1}\ln\frac{\ln(\mu/\Lambda)}{\ln(1/(b\Lambda))} 
\ee
In practical calculations, $\mu$ is identified with the factorization
scale of the hard kernel. It is to be noted that for the B-meson, the
Sudakov factor is considered only for the lighter quark. Therefore,
the exponents of the individual Sudakov factors 
(including the evolution function)
are given as follows:
\be
S_{\rho} = s(x,\brho,\mb) + s(1-x,\brho,\mb) - 
\frac{1}{\beta_1}\ln\frac{\ln(\mu/\Lambda)}{\ln(1/(\brho\Lambda))}
\ee
\be
S_B = s(\xi,\bb,\mb) - 
\frac{1}{\beta_1}\ln\frac{\ln(\mu/\Lambda)}{\ln(1/(\bb\Lambda))}
\ee
\indent The last major ingredient to be introduced is the threshold
resummation. The double logarithms of the form $\ln^2x$ diverge at the
endpoints and therefore they are also resummed. In order to achieve
this goal, a jet function, $S_t(x)$ is introduced which vanishes at
the endpoints. It has been proposed \cite{sanda}
that for phenomenological studies,
a simple parameterization can be used. The proposed parameterization is
\be
S_t(x) = \frac{2^{1+2c}\Gamma(3/2+c)}{\sqrt{\pi}\Gamma(1+c)}[x(1-x)]^c
\ee
with $c\sim 0.3$. We use this parameterization in our numerical study
but agree that this may not be the accurate form.\\
\indent Finally, we remark that it is not very clear whether the power
suppressed corrections arising due to working in different gauges for
different quantities are going to play an important role
\cite{sachrajda}.
This is one
feature that has to be checked but we leave it for a separate study.
The DAs discussed in the previous section are assumed to be finally
multiplied by the corresponding Sudakov factors, the transverse
dependence carrying functions - the $\Sigma$'s and threshold functions.
\end{section}
\begin{section}{Evaluating $\btorho$ form factors}
\indent The $\btorho$ form factors as introduced above can also be
written in a slightly different form. We make the following
identifications. The coefficients of $\pb+\prho$ and $\pb-\prho$
define the form factors $A_2$ and $A_3$, as discussed above. Instead,
we define ${\tilde{A}}_2$ and ${\tilde{A}}_2$ as coefficients of $\pb$
and $\prho$ such that $A_2$ and $A_3$ are simply sum and difference of
the new form factors. The remaining two have the same definition as
above. This has been done for convenience only.\\
\indent We evaluate the ``parallel'' and ``transverse'' contributions
from both the diagrams separately. The individual expressions read
(again as mentioned in the last section, we assume the DAs to be
multiplied by suitable Sudakov, threshold and momentum dependence
factors and do not write them explicitly here in the expressions)
\bea
{\mathcal{M}}^{(1)}_{\mu\parallel} &=& (-4\pi N_cC_F) 
\Bigg(\frac{i\fb\mrho^2\mb^2}
{16\sqrt{2}E}\Bigg)(\ep\cdot\nplus)\int dx~d\xi ~d^2{\bf \lperp}~
d^2{\bf \kperp}~
\alpha_s \\ \nonumber &\Bigg\{&{\prho}_{\mu}
\Bigg(\frac{2iE\frho}{\mrho}\phiparallel
\Bigg[\frac{\sqrt{2}\eta}{\mb(\eta^2-r^2)}\psibbar
  + \frac{4\eta}{\mb(\eta^2-r^2)}\psib \\ \nonumber
&+& \frac{\sqrt{2}\eta^2}{\mb(\eta^2-r^2)}
   +  \frac{2x}{\mb}\psib + \frac{\sqrt{2}x}{\mb}\psibbar + 
\frac{2x(\eta +r^2)\eta}{\mb(\eta^2-r^2)}\psib \Bigg] \\ \nonumber
&+& \fperp\hparallelt\Bigg[\frac{2\sqrt{2}i\eta}{\mb(\eta^2-r^2)}\psib +
\frac{2i\eta}{\mb(\eta^2-r^2)}\psibbar -
\frac{4\sqrt{2}i\eta^2x}{\mb(\eta^2-r^2)}\psib\Bigg] \\ \nonumber
&+& 4i\fperp
E\Bigg[\int_0^xdv[\phi_{\perp}(v)-h^{(t)}_{\parallel}(v)]\Bigg]
\deltal\frac{\eta}{\mb(\eta^2-r^2)}~(\delkperp\cdot\dellperp) \\ \nonumber
&+&
2\sqrt{2}i\fperp\hparallelsprime\Bigg[-\frac{\eta}
{\sqrt{2}\mb(\eta^2-r^2)}\psibbar + \frac{2x}{\mb}\psib\Bigg]\Bigg) \\
\nonumber \\ \nonumber
&\bigoplus& {\pb}_{\mu}\Bigg(\frac{2iE\frho}{\mrho}\phiparallel\Bigg[
\frac{\sqrt{2}\eta^2}{\mb(\eta^2-r^2)}\psibbar -
\frac{\sqrt{2}\eta^2r^2x}{\mb(\eta^2-r^2)}\psibbar -
\frac{\sqrt{2}}{\mb}\psibbar - \frac{2r^2x}{\mb}\psib\Bigg] \\
\nonumber
&+&
\fperp\hparallelt\Bigg[-\frac{\sqrt{2}i\eta^2}{\mb(\eta^2-r^2)}\psibbar
  + \frac{2i\eta^2}{\mb(\eta^2-r^2)}\psibbar +
  \frac{4\sqrt{2}i\eta^2r^2x}{\mb(\eta^2-r^2)}\psib\Bigg] \\ \nonumber
&+&
2\sqrt{2}i\fperp\hparallelsprime\Bigg[\frac{\eta^2}
{\sqrt{2}\mb(\eta^2-r^2)}\psibbar - \frac{1}{\mb}\psib\Bigg]
\Bigg)\Bigg\}
\otimes \frac{1}{[x\xi\eta\mb^2 -
    (\lperp-\kperp)^2][x\eta\mb^2-\kperp^2]} 
\eea
where the derivatives should be viewed as acting on the hard kernel
and we have already set terms proportional to $\kperp$ and $\lperp$
appearing in the numerator to be zero. Similarly we have
\bea
{\mathcal{M}}^{(2)}_{\mu\parallel} &=& (-4\pi N_cC_F) 
\Bigg(\frac{i\fb\mrho^2\mb^2}
{8E}\Bigg)(\ep\cdot\nplus)\int dx~d\xi~ d^2{\bf \lperp}~d^2{\bf \kperp}~
\alpha_s \\ \nonumber &\Bigg\{&{\prho}_{\mu}
\Bigg(\frac{2iE\frho}{\mrho}\phiparallel
\Bigg[-\frac{\sqrt{2}\xi\eta^2}{\mb(\eta^2-r^2)}\psib -
  \frac{\xi\eta}{\mb(\eta^2-r^2)}\psibbar +
  \frac{\eta^2}{\mb(\eta^2-r^2)}\psibbar \\ \nonumber
&-& \frac{1}{\mb}\psibbar -
  \frac{1}{\sqrt{2}\mb}\psib +
  \frac{1}{\sqrt{2}}\frac{\eta(\eta+r^2)}{\mb(\eta^2-r^2)}\psib\Bigg]
\\ \nonumber
&+&
2i\fperp\hparallelsprime\Bigg[\frac{\sqrt{2}\xi\eta}{\mb(\eta^2-r^2)}\psibbar
  - \frac{1}{\mb}\psib -
  \frac{1}{\sqrt{2}}\frac{\eta(\eta+r^2)}{\mb(\eta^2-r^2)}\psibbar\Bigg]\Bigg)
 \\ \nonumber \\ \nonumber
&\bigoplus&
    {\pb}_{\mu}\Bigg(\frac{2iE\frho}{\mrho}\phiparallel
\Bigg[-\frac{r^2\eta^2}{\mb(\eta^2-r^2)}\psibbar - 
\frac{r^2}{\sqrt{2}\mb}\psib +
\frac{\xi\eta^2}{\mb(\eta^2-r^2)}\psibbar + 
\frac{\sqrt{2}\xi}{\mb}\psib\Bigg] \\ \nonumber
&+& 2i\fperp\hparallelsprime\Bigg[\frac{r^2}{\sqrt{2}\mb}\psibbar -
      \frac{\sqrt{2}\xi\eta^2}{\mb(\eta^2-r^2)}\psibbar +
      \frac{\xi}{\mb}\psib  \\ \nonumber
&-&\frac{\eta}{\sqrt{2}\mb}\psibbar + 
\frac{1}{\sqrt{2}}\frac{\eta^2(\eta+r^2)}{\mb(\eta^2-r^2)}\psibbar\Bigg]
\Bigg)\Bigg\}
\otimes \frac{1}{[x\xi\eta\mb^2 -
    (\lperp-\kperp)^2][\xi\eta\mb^2-\lperp^2]} 
\eea
\\
\bea
{\mathcal{M}}^{(1)}_{\mu\perp} &=& (-4\pi
N_cC_F)\Bigg(-\frac{\fb\mb^2}{4}\Bigg)\int dx~d\xi~ d^2{\bf
  \lperp}~d^2{\bf \kperp}~\alpha_s \\ \nonumber
&\Bigg\{&i\epsilon_{\mu\nu\alpha\beta}\pb^{\nu}\prho^{\alpha}{\epperp}^{\beta}
\Bigg(\fperp E\phiperp\Bigg[\frac{2xr^2\eta}{\mb^2(\eta^2-r^2)}\psibbar
  - \frac{2x\eta^2}{\mb^2(\eta^2-r^2)}\psibbar -
  \frac{2\sqrt{2}\eta}{\mb^2(\eta^2-r^2)}\psib \\ \nonumber
&+& \frac{2x}{\mb^2}\psibbar -
  \frac{2\eta^3}{\mb^2(\eta^2-r^2)}\psibbar -
  \frac{2x\eta^2}{\mb^2(\eta^2-r^2)}\psibbar +
  \frac{2x\eta^3(\eta+r^2)}{\mb^2(\eta^2-r^2)}\psibbar\Bigg] \\
\nonumber
&+& \frho\mrho\gperpv\Bigg[\frac{2x}{\mb^2}\psib -
  \frac{\sqrt{2}\eta}{\mb^2(\eta^2-r^2)}\psibbar + 
\frac{\sqrt{2}\eta^2}{\mb^2(\eta^2-r^2)}\psibbar \Bigg]\\ \nonumber
&+&
\frac{\frho\mrho\gperpaprime}{8}\Bigg[\frac{\sqrt{2}xr^2\eta^3}
{\mb^2(\eta^2-r^2)^2}\psibbar -
\frac{4\eta^3}{\mb^2(\eta^2-r^2)^2}\psib -
\frac{\sqrt{2}x\eta^4}{\mb^2(\eta^2-r^2)^2}\psibbar -
\frac{x\eta^3(\eta+r^2)}{\mb^2(\eta^2-r^2)^2}\psib\Bigg]\Bigg) \\
\nonumber \\ \nonumber 
&\bigoplus&
(\ep\cdot\pb){\pb}_{\mu}\Bigg(\frho\mrho\gperpv\Bigg[
\frac{\sqrt{2}\eta^2}{\mb^2(\eta^2-r^2)^2}\fac\psibbar
  - \frac{\eta^2}{\sqrt{2}\mb^2(\eta^2-r^2)^2}\psibbar\Bigg] \\
\nonumber
&-&
\frac{\frho\mrho\gperpaprime}{8}\Bigg[\frac{\sqrt{2}x\eta^4r^2}
{\mb^2(\eta^2-r^2)^2}\psibbar +
\frac{2xr^2\eta^2}{\mb^2(\eta^2-r^2)^2}\fac\psib \\ \nonumber &-&
\frac{\eta^4}{\sqrt{2}\mb^2(\eta^2-r^2)^2}\psibbar +
\frac{\sqrt{2}\eta^2}{\mb^2(\eta^2-r^2)^2}\fac\psibbar\Bigg]\Bigg) \\
\nonumber \\ \nonumber
&\bigoplus& \ep_{\mu}\Bigg(\fperp E\phiperp\Bigg[-2\sqrt{2}xr^2\psib -
  \psibbar + \sqrt{2}\psib\Bigg] \\ \nonumber 
&+& \frho\mrho\gperpv\Bigg[-\frac{xr^2}{\sqrt{2}}\psibbar + 2\psib +
  \frac{x\eta}{\sqrt{2}}\psibbar + x(\eta+r^2)\psib\Bigg] \\ \nonumber
&-&
\frac{\frho\mrho\gperpaprime}{8}\Bigg[\frac{xr^2}{\sqrt{2}}\psibbar -
  xr^2\psib + \frac{x\eta}{\sqrt{2}}\psibbar + x\eta\psib -
  \sqrt{2}\psibbar\Bigg] \\ \nonumber
&+& \frac{\frho\mrho
  E\gperpa}{4}\Bigg[\frac{xr^2}{\sqrt{2}}\deltal(\delkperp\cdot\dellperp) - 
\frac{1}{\sqrt{2}}\deltal(\delkperp\cdot\dellperp)\Bigg]\Bigg) \\ \nonumber \\ \nonumber
&\bigoplus& (\ep\cdot\prho){\prho}_{\mu}\Bigg(\fperp E\phiperp\Bigg[  
\frac{4\sqrt{2}xr^2\eta^3}{\mb^2(\eta^2-r^2)^2}\psib +
\frac{2\eta^3}{\mb^2(\eta^2-r^2)^2}\psibbar -
\frac{2\sqrt{2}\eta^3}{\mb^2(\eta^2-r^2)^2}\psib \\ \nonumber
&+& \frac{2\eta^3}{\mb^2(\eta^2-r^2)^2}\psibbar -
\frac{2\sqrt{2}\eta}{\mb^2(\eta^2-r^2)^2}\fac\psib\Bigg] \\ \nonumber
&+&
\frho\mrho\gperpv\Bigg[\frac{2xr^2\eta^3}{\mb^2(\eta^2-r^2)^2}\psibbar
  - \frac{4\eta^3}{\mb^2(\eta^2-r^2)^2}\psib -
  \frac{\sqrt{2}x\eta^4}{\mb^2(\eta^2-r^2)^2}\psibbar -
  \frac{2x\eta^3(\eta+r^2)}{\mb^2(\eta^2-r^2)^2}\psib \\ \nonumber
&+& \frac{\sqrt{2}x\eta^2}{\mb^2(\eta^2-r^2)^2}\psibbar -
  \frac{\sqrt{2}\eta}{\mb^2(\eta^2-r^2)^2}\fac\psibbar +
  \frac{2x}{\mb^2}\fac\psib\Bigg] \\ \nonumber
&-&
\frac{\frho\mrho\gperpaprime}{8}\Bigg[-\frac{\sqrt{2}xr^2\eta^3}
{\mb^2(\eta^2-r^2)^2}\psibbar +
\frac{2xr^2\eta^3}{\mb^2(\eta^2-r^2)^2}\psib -
\frac{\sqrt{2}x\eta^4}{\mb^2(\eta^2-r^2)^2}\psibbar -
\frac{2x\eta^4}{\mb^2(\eta^2-r^2)^2}\psib \\ \nonumber
&+& \frac{2\sqrt{2}\eta^3}{\mb^2(\eta^2-r^2)^2}\psibbar -
\frac{x\eta4}{\sqrt{2}\mb^2(\eta^2-r^2)^2}\psibbar +
\frac{\eta3}{\sqrt{2}\mb^2(\eta^2-r^2)^2}\psibbar \\ \nonumber
&-& \frac{2x\eta^2}{\mb^2(\eta^2-r^2)^2}\fac\psibbar +
\frac{\sqrt{2}\eta}{\mb^2(\eta^2-r^2)^2}\fac\psibbar\Bigg] \\
\nonumber
&+& \frac{\frho\mrho
  E\gperpa}{4}\Bigg[\frac{\sqrt{2}\eta}{\mb^2(\eta^2-r^2)^2}
\deltal\fac(\delkperp\cdot\dellperp) 
\Bigg]\Bigg)\Bigg\}
\otimes \frac{1}{[x\xi\eta\mb^2 -
    (\lperp-\kperp)^2][x\eta\mb^2-\kperp^2]} 
\eea
\\
\bea
{\mathcal{M}}^{(2)}_{\mu\perp} &=& (-4\pi
N_cC_F)\Bigg(-\frac{\fb\mb^2}{4}\Bigg)\int dx~d\xi~ d^2{\bf
  \lperp}~d^2{\bf \kperp}~\alpha_s \\ \nonumber
&\Bigg\{&i\epsilon_{\mu\nu\alpha\beta}\pb^{\nu}\prho^{\alpha}{\epperp}^{\beta}
\Bigg(\frho\mrho\gperpv\Bigg[-\frac{2i}{\mb^2}\psib
  + \frac{\sqrt{2}i\xi\eta}{\mb^2(\eta^2-r^2)}\psibbar -
  \frac{\sqrt{2}i\eta^2}{\mb^2(\eta^2-r^2)}\psibbar\Bigg] \\ \nonumber
&+&
\frac{\frho\mrho\gperpaprime}{8}\Bigg[\frac{\sqrt{2}r^2\eta^3}
{\mb^2(\eta^2-r^2)^2}\psibbar +
\frac{4\xi\eta^3}{\mb^2(\eta^2-r^2)^2}\psib -
\frac{\sqrt{2}\eta^4}{\mb^2(\eta^2-r^2)^2}\psibbar -
\frac{2\eta^3(\eta+r^2)}{\mb^2(\eta^2-r^2)^2}\psib\Bigg]\Bigg) \\
\nonumber \\ \nonumber
&\bigoplus&
(\ep\cdot\pb){\pb}_{\mu}\Bigg(\frho\mrho\gperpv\Bigg[\frac{\xi\eta^2}{\sqrt{2}\mb^2(\eta^2-r^2)}\psibbar
  + \frac{\xi\eta^2}{\mb^2(\eta^2-r^2)}\fac\psibbar \\ \nonumber &+&
  \frac{4\xi}{\mb^2}\fac\psib\Bigg]
+
  \frac{\frho\mrho\gperpaprime}{8}\Bigg[\frac{r^2\eta^4}
{\sqrt{2}\mb^2(\eta^2-r^2)^2}\psibbar + 
\frac{2r^2}{\mb^2(\eta^2-r^2)}\fac\psib \\ \nonumber &-& \frac{\xi\eta^4}
{\sqrt{2}\mb^2(\eta^2-r^2)^2}\psibbar + \frac{\xi}{\mb^2(\eta^2-r^2)}
\fac\psibbar\Bigg]\Bigg) \\ \nonumber \\ \nonumber
&\bigoplus&
       {\ep}_{\mu}\Bigg(\frho\mrho\gperpv\Bigg[-\frac{r^2}{\sqrt{2}}\psibbar 
- 2\xi\psib + \frac{\eta}{\sqrt{2}}\psibbar +(\eta+r^2)\psib\Bigg] \\
       \nonumber
&+& \frac{\frho\mrho\gperpaprime}{8}\Bigg[\frac{r^2}{\sqrt{2}}\psibbar
	 - r^2\psib - \sqrt{2}\xi\psibbar +
	 \frac{\eta}{\sqrt{2}}\psibbar + \eta\psib\Bigg] \\ \nonumber
&+& \frac{\frho\mrho E\gperpa}{4}\Bigg[\frac{\xi}{\sqrt{2}}
\deltal(\delkperp\cdot\dellperp) - \frac{r2}{\sqrt{2}}
\deltal(\delkperp\cdot\dellperp)\Bigg]\Bigg) \\ \nonumber \\ \nonumber
&\bigoplus& (\ep\cdot\pb){\prho}_{\mu}\Bigg(\frho\mrho\gperpv\Bigg[
\frac{\sqrt{2}r^2\eta^3}{\mb^2(\eta^2-r^2)^2}\psibbar +
  \frac{4\xi\eta^3}{\mb^2(\eta^2-r^2)^2}\psib \\ \nonumber &-&
  \frac{\sqrt{2}\eta^4}{\mb^2(\eta^2-r^2)^2}\psibbar -
  \frac{2\eta^3(\eta+r^2)}{\mb^2(\eta^2-r^2)^2}\psib 
- \frac{\eta^2}{\sqrt{2}\mb^2(\eta^2-r^2)}\psib \\ \nonumber &-&
  \frac{\sqrt{2}\xi\eta}{\mb^2(\eta^2-r^2)}\fac\psibbar -
  \frac{2}{\mb^2}\fac\psib\Bigg] \\ \nonumber
&+&
\frac{\frho\mrho\gperpaprime}{8}\Bigg[-\frac{\sqrt{2}r^2\eta^3}
{\mb^2(\eta^2-r^2)^2}\psibbar
  + \frac{2r^2\eta^3}{\mb^2(\eta^2-r^2)^2}\psib +
  \frac{2\sqrt{2}\xi\eta^3}{\mb^2(\eta^2-r^2)^2}\psibbar -
  \frac{2r^2\eta^4}{\mb^2(\eta^2-r^2)^2}\psibbar \\ \nonumber
&-& \frac{\sqrt{2}\eta^4}{\mb^2(\eta^2-r^2)^2}\psib +
  \frac{\xi\eta^3}{\sqrt{2}\mb^2(\eta^2-r^2)^2}\psibbar -
  \frac{\eta^4}{\sqrt{2}\mb^2(\eta^2-r^2)^2}\psibbar \\ \nonumber
&+& \frac{\xi\eta}{\mb^2(\eta^2-r^2)}\fac\psibbar -
  \frac{\sqrt{2}\eta^2}{\mb^2(\eta^2-r^2)}\fac\psib\Bigg] \\ \nonumber
&+& \frac{\frho\mrho E\gperpa}{4}\Bigg[\frac{\sqrt{2}\xi\eta^3}
{\mb^2(\eta^2-r^2)^2}\deltal(\delkperp\cdot\dellperp) + 
\frac{\sqrt{2}r^2\eta^3}{\mb^2(\eta^2-r^2)}\deltal(\delkperp\cdot\dellperp)
\\ \nonumber
&-& \frac{\xi\eta}{\mb^2(\eta^2-r^2)}\fac\deltal(\delkperp\cdot\dellperp)
\Bigg]\Bigg)\Bigg\}
\otimes \frac{1}{[x\xi\eta\mb^2 -
    (\lperp-\kperp)^2][\xi\eta\mb^2-\lperp^2]}
\eea
These expressions are Fourier transformed to the $b$-space. 
The wave functions/DAs are assumed to have been multiplied by the
Sudakov, threshold resummation and momentum dependence factors. While
Fourier transforming the expressions, we introduce the following $h_i$
functions:
\be
h_1 = K_0(\sqrt{x\xi\eta}\mb\bb)\{\theta(\bb-\brho)I_0(\sqrt{x\eta}\mb\brho)
K_0(\sqrt{x\eta}\mb\bb)
 + \theta(\brho-\bb)I_0(\sqrt{x\eta}\mb\bb)K_0(\sqrt{x\eta}\mb\brho)\}
\ee
\be
h_2 = K_0(\sqrt{x\xi\eta}\mb\brho)\{\theta(\bb-\brho)
I_0(\sqrt{\xi\eta}\mb\brho)
K_0(\sqrt{\xi\eta}\mb\bb)
 + \theta(\brho-\bb)I_0(\sqrt{\xi\eta}\mb\bb)K_0(\sqrt{\xi\eta}\mb\brho)\}
\ee 
\be
h_3 = K_1(\sqrt{x\xi\eta}\mb\bb)\{\theta(\bb-\brho)I_0(\sqrt{x\eta}\mb\brho)
K_0(\sqrt{x\eta}\mb\bb)
 + \theta(\brho-\bb)I_0(\sqrt{x\eta}\mb\bb)K_0(\sqrt{x\eta}\mb\brho)\}
\ee
\be
h_4 = K_1(\sqrt{x\xi\eta}\mb\brho)\{\theta(\bb-\brho)
I_0(\sqrt{\xi\eta}\mb\brho)
K_0(\sqrt{\xi\eta}\mb\bb)
 + \theta(\brho-\bb)I_0(\sqrt{\xi\eta}\mb\bb)K_0(\sqrt{\xi\eta}\mb\brho)\}
\ee 
and finally set $r^2=0$ to retain terms up to twist-3.\\
\indent Before ending this section we would like to discuss the
singular nature of the strong coupling constant as the
scale $\Lambda$ is approached. Close to this scale, it is easy to
convince oneself that the perturbation theory should break down and
therefore the calculation outlined above makes no sense as this scale
is approached. For the reliability of any perturbative calculation, 
it has to be ensured that most of the contribution to the calculated
quantity comes from the perturbative regime. It has been observed for
the pion form factor that this is not really true and a large
contribution does come from the non-perturbative region, something
which makes the predictions highly unreliable. Near this scale, the
coupling constant acquires a large value, thereby, rendering the
breakdown of the RG improved perturbation theory. The infra-red
behaviour of the strong coupling constant is one of the major
challenges of modern day physics. However, it has been observed that
there is a good phenomenological evidence to believe that close to
(and below) the scale $\Lambda$, the coupling constant gets {\it
  frozen} to a value which is not a big number literally (see
\cite{frozen} and references therein). The concept
of a frozen coupling constant was proposed long ago by
Cornwall \cite{cornwall}. Also, it has been proposed that the gluon
propagator should
be modified in order to circumvent the problem of
singularity in the coupling constant and this has proved a successful
phenomenological ansatz. If one adopts this point of view that as one
approaches the scale $\Lambda$, the gluon propagator should be
appropriately modified (using some regulator, say) and if one also assumes
that the coupling constant gets frozen to some fixed value, then, one
can try to carry out the analysis with the frozen value of the
coupling constant for scales close to or below $\Lambda$. If the
contributions are small, then this establishes the validity of the
perturbative treatment.    
\end{section}
\begin{section}{Behaviour of form factors and rough numerical estimates}
\indent The various form factors are read from the expressions
obtained above.
To numerically study the behaviour of the form factors, the choice of
the scale $\mu$ has to be made. We take $\mu =
Max(\sqrt{x\xi\eta}\mb,1/\bb,1/\brho)$. This scale must be greater
than $\Lambda$ in order to avoid the singular behaviour of the strong
coupling constant. We would like to emphasize again that in this study
we do not make any attempt of precisely calculating the form factors
but focus on the behaviour of the same as functions of $\bb$ and
$\brho$ and see whether the perturbative calculations make any sense.
The choices for the other parameters are as
follows:
\[
\Lambda = 0.25~ GeV \hskip 0.5cm \mb = 5.279 ~GeV \hskip 0.5cm  
\mrho = 0.77 ~GeV
\]
\[
\fb = 0.18 ~GeV \hskip 0.5cm \frho = 0.198 ~GeV \hskip 0.5cm 
\fperp = 0.152 ~GeV
\]
\[
\wb = 0.35 ~GeV \hskip 1cm \wrho = 0.3 ~GeV
\]
The parameters $\wb$ and $\wrho$ should be ${\mathcal{O}}(\Lambda)$
and therefore have been chosen as above.
\indent Figures 2 to 5 show the behaviour of various
form factors for large recoil ie. $\qsq\to 0$ or $\eta\to 1$. From the
figures it is quite evident that the Sudakov and threshold resummation
factors are successful in regulating the endpoint behaviour of the
form factors. Furthermore, it is not hard to convince oneself by
looking at the figures that the perturbative region supplies the
dominant contribution. Also, it can be seen that for intermediate
values of $\bb$ and $\brho$, there is indeed an enhancement as can be
expected and as discussed earlier. The Sudakov suppression is much
weaker for the B-meson as is clear from the variation with $\bb$. The
reason for the appearance of a turning point while studying the
variation with $\bb$ is the fact that the Sudakov suppression is
weaker and there is a tendency for the strong coupling constant to
blow up as $1/\bb\to \Lambda$. However, if one assumes the
phenomenological ansatz of frozen coupling constant as $\Lambda$ is
approached, then explicit evaluations show that the contributions from
the non-perturbative region are small.\\
\indent We summarize the numerical values for various form factors
with these choices of parameters:
\[
V = 0.43 \hskip 1.5cm A_1 = 0.39 \]
\[ A_2 = 1.73 \hskip 1.5cm A_3 = 1.92 \]
Very evidently, $V$ and $A_1$ are slightly overestimated while $A_2$
and $A_3$ deviate significantly \cite{ball3}. 
We would like to stress upon the fact
that in this particular study we have not bothered to carry out a
precise numerical analysis which would involve varying various other
parameters like $\wb$ and $\wrho$ and then finally concluding which
choices give the best results. It is straightforward to note that
slightly higher values for both these parameters would have given more
suppression. This has been the case with \cite{sanda} and  \cite{wei}.
The best value which the authors in \cite{sanda}
use for $\wb$ was obtained
by comparing the pQCD results to the pion form factor with the use of
single wave function for the B-meson. However, when the other pieces
are also included, this value is bound to change.\\
\indent In \cite{wei}, it has been remarked that the contribution
from the $\deltal$ term is roughly $20\%$. We agree with the general
arguments and conclusions and reiterate that this term is
crucial. Also, the $\psibbar$ term, though is sub-leading, but can
still change the results when incorporated by at least a few
percent. Furthermore, setting the two B-meson wave functions same is
equivalent to setting the $\psibbar$ term to zero and also the
$\deltal$ term to zero. Therefore, the use of a single wave function
for the B-meson does more than expected and therefore introduces more
uncertainty than thought of.
We attribute the reason of large deviations in $A_2$ and $A_3$ to the
presence of not really insignificant $\psibbar$ and $\deltal$
terms. However, as mentioned earlier, the power suppressed terms that
would arise due to different choices of gauges for defining 
different quantities have not been included and the relative sign of
the same can be crucial in a complete and accurate numerical analysis.
\end{section}
\begin{section}{Conclusions and summary}
\indent We have studied the $\btorho$ form factors within the
perturbative QCD approach including the twist-3 contributions to the
DAs. For the B-meson, we employed a model for the wave
functions that satisfies the relevant equations of motion and passes
through this level of criticism. The use of single wave function has
been criticised and we avoid this aspect by retaining the so called
sub-leading pieces, which turn out to be not so sub-leading and
insignificant. The appropriate Sudakov and threshold
resummation functions have been included. With these ingredients the
results in figures clearly are very encouraging and seem to suggest
that the pQCD calculation of the various form factors is reliable and
valid in the sense that dominant contribution clearly appears to stem
from the perturbative regime. However, since the aim was not a very
accurate determination of the numerical values, we have overestimated,
at least two of the form factors. But we believe that a careful choice
of various parameters can lead to a consistent determination of the
same.
Also, there can be slight differences once a complete and precise 
jet function is employed in the calculations rather than the
phenomenologically motivated function that has been used here.\\
\indent Encouraged by the results obtained, and mainly the fact that
the endpoint behaviour is actually regulated and the contribution from
the perturbative regime is the dominant one, we would like to propose
the following as a working hypothesis for such calculations:
\begin{enumerate}
\item[$*$] Use of two wave functions for the B-meson ie. retaining the
  generally thought sub-leading terms.
\item[$*$] Including proper Sudakov and threshold resummation
\item[$*$] To use, as a phenomenological ansatz, the notion of a
  frozen coupling constant for scales very near or below
  $\Lambda_{QCD}$.   
\end{enumerate} 
The precise form of the B-meson wave functions is not known but we
believe that the results will only improve with the advancement of our
understanding of them. The results of this study, particularly the
behaviour of the form factors with $\bb$ and $\brho$ are very
encouraging and require a detailed and careful numerical scrutiny.  
\end{section}

\pagebreak
\vskip 2.0cm
\begin{figure}[ht]
\centerline{
\epsfxsize=7.0cm\epsfysize=6.0cm
                      \epsfbox{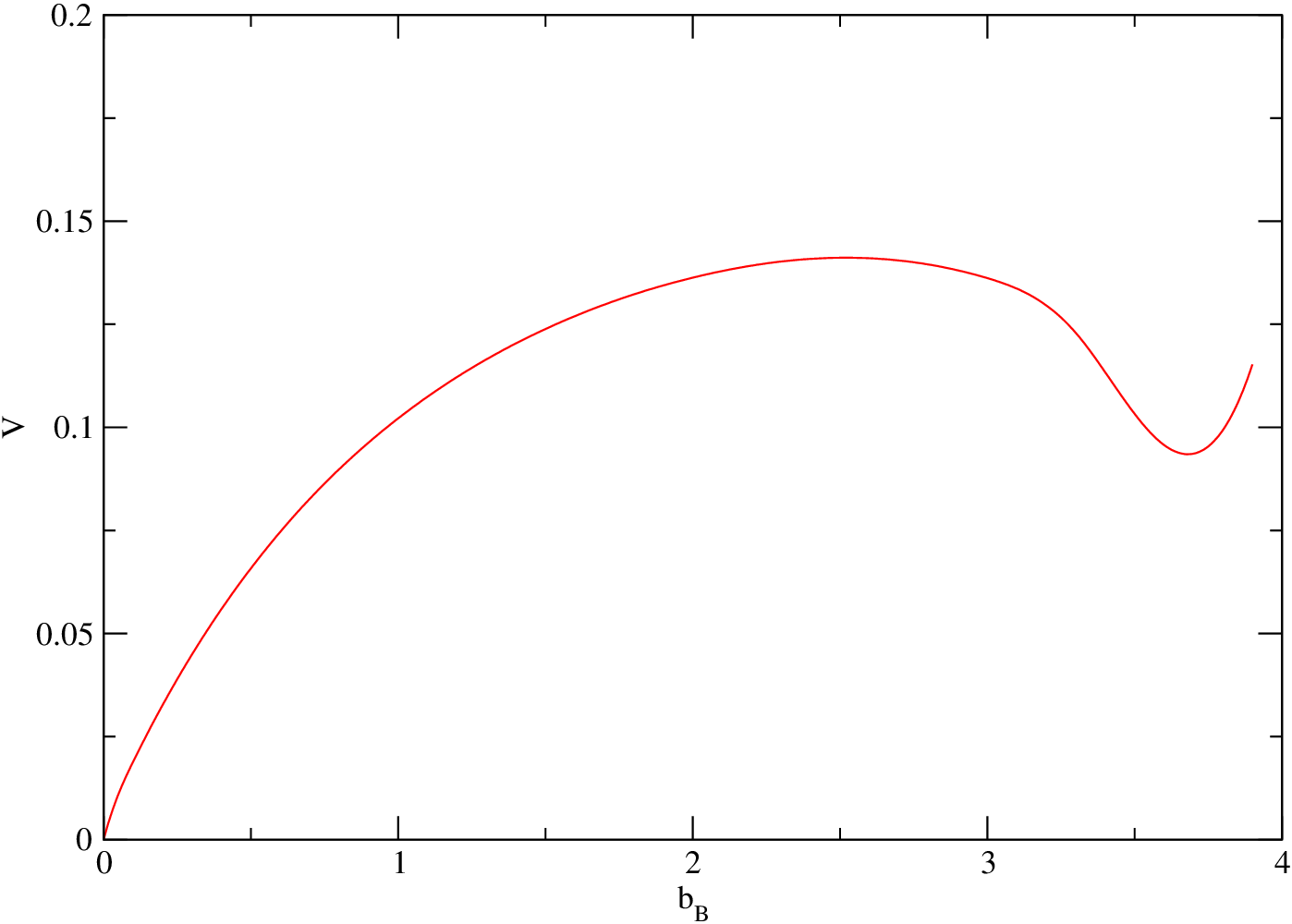}
\hskip 1.0cm 
\epsfxsize=7.0cm\epsfysize=6.0cm
                      \epsfbox{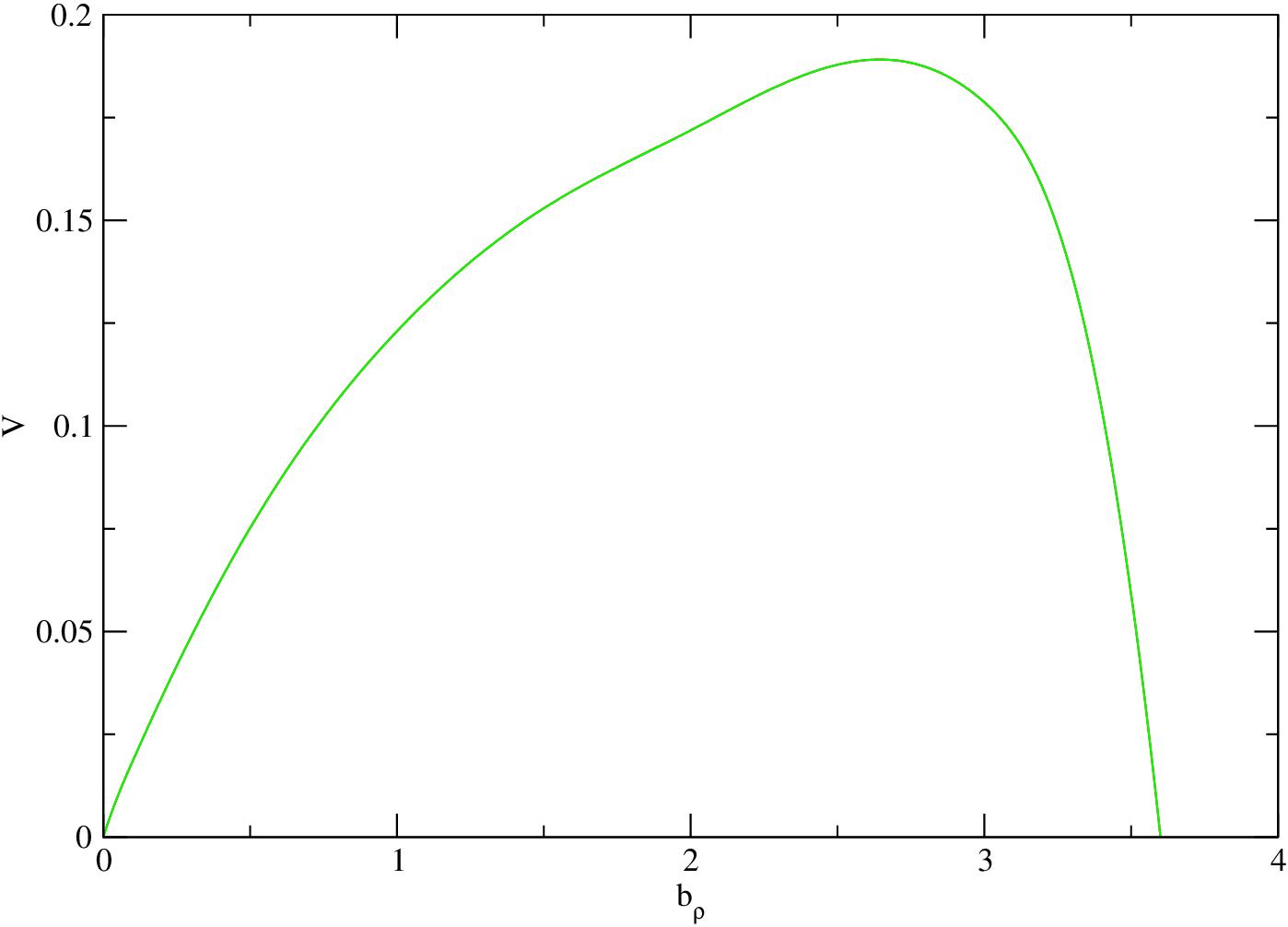}}
\caption{Variation of $V$ with $\bb$ (left) and $\brho$ (right).}
\end{figure}
\vspace*{6.0cm}
\begin{figure}[ht]
\centerline{
\epsfxsize=7.0cm\epsfysize=6.0cm
                      \epsfbox{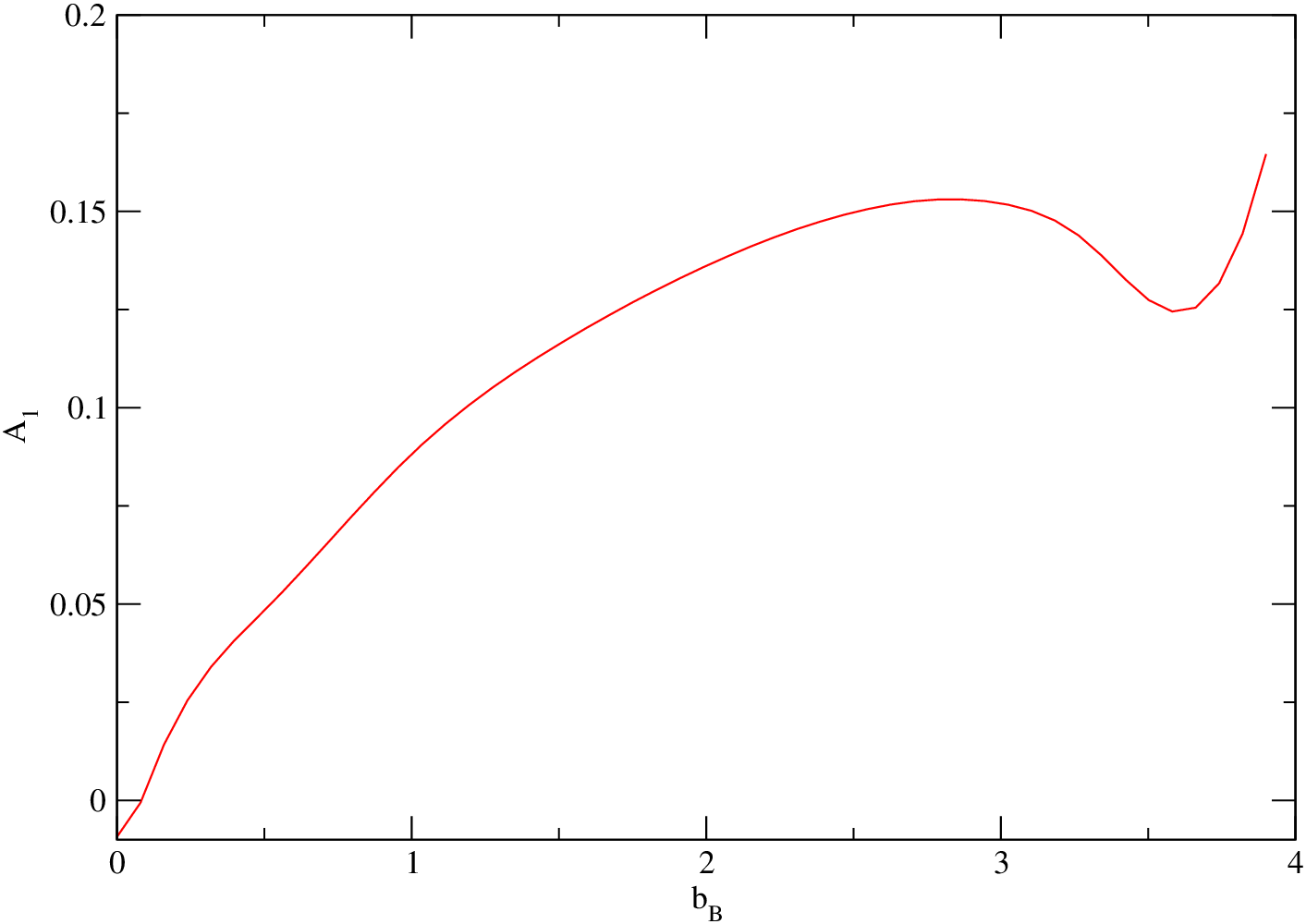}
\hskip 1.0cm 
\epsfxsize=7.0cm\epsfysize=6.0cm
                      \epsfbox{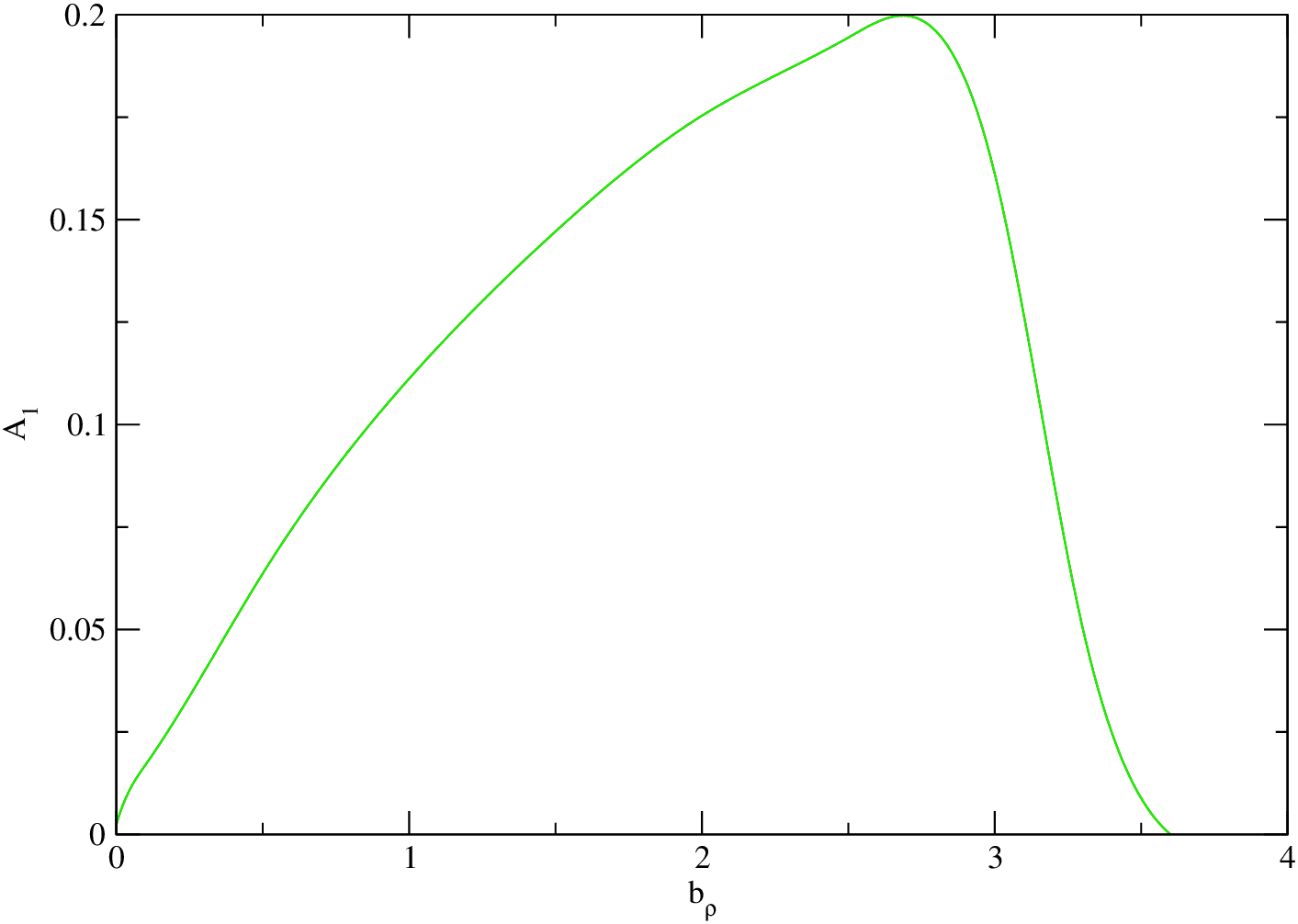}}
\caption{Variation of $A_1$ with $\bb$ (left) and $\brho$ (right).\\}
\end{figure}
\vskip 2.0cm
\begin{figure}[ht]
\centerline{
\epsfxsize=7.0cm\epsfysize=6.0cm
                      \epsfbox{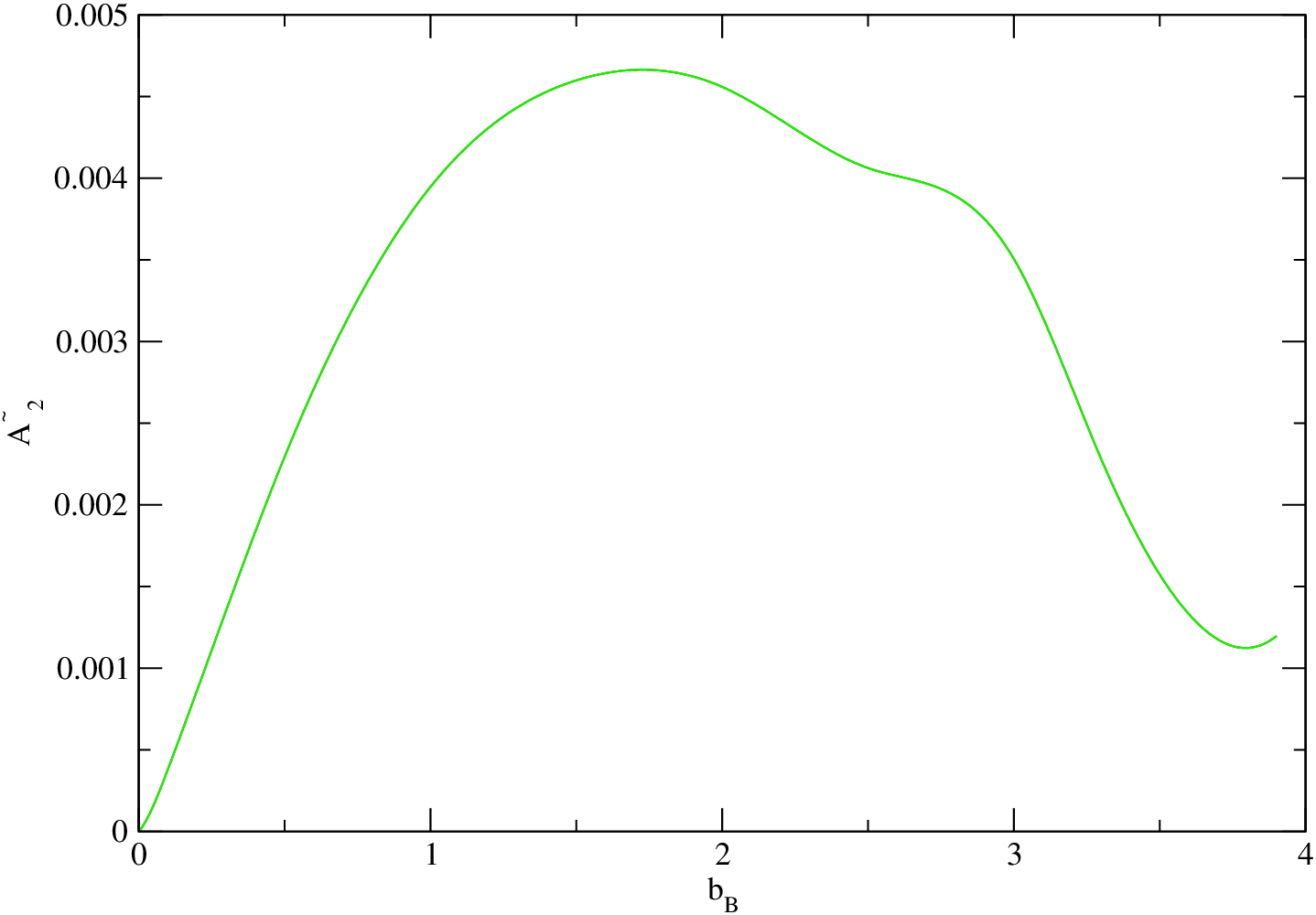}
\hskip 1.0cm 
\epsfxsize=7.0cm\epsfysize=6.0cm
                      \epsfbox{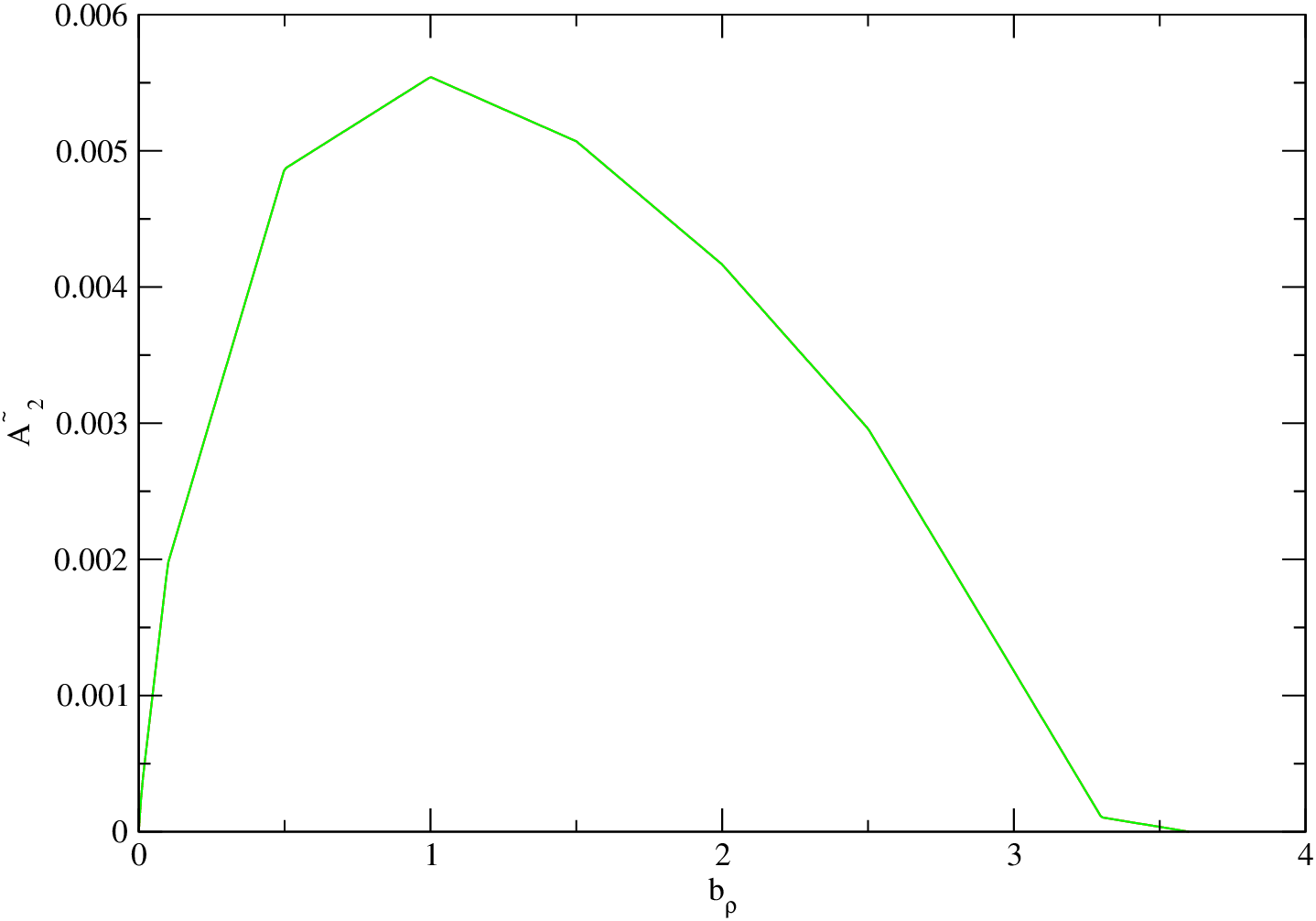}}
\caption{Variation of ${\tilde{A}}_2$ with $\bb$ (left) and $\brho$ (right).\\}
\end{figure}
\vspace*{6cm}
\begin{figure}[ht]
\centerline{
\epsfxsize=7.0cm\epsfysize=6.0cm
                      \epsfbox{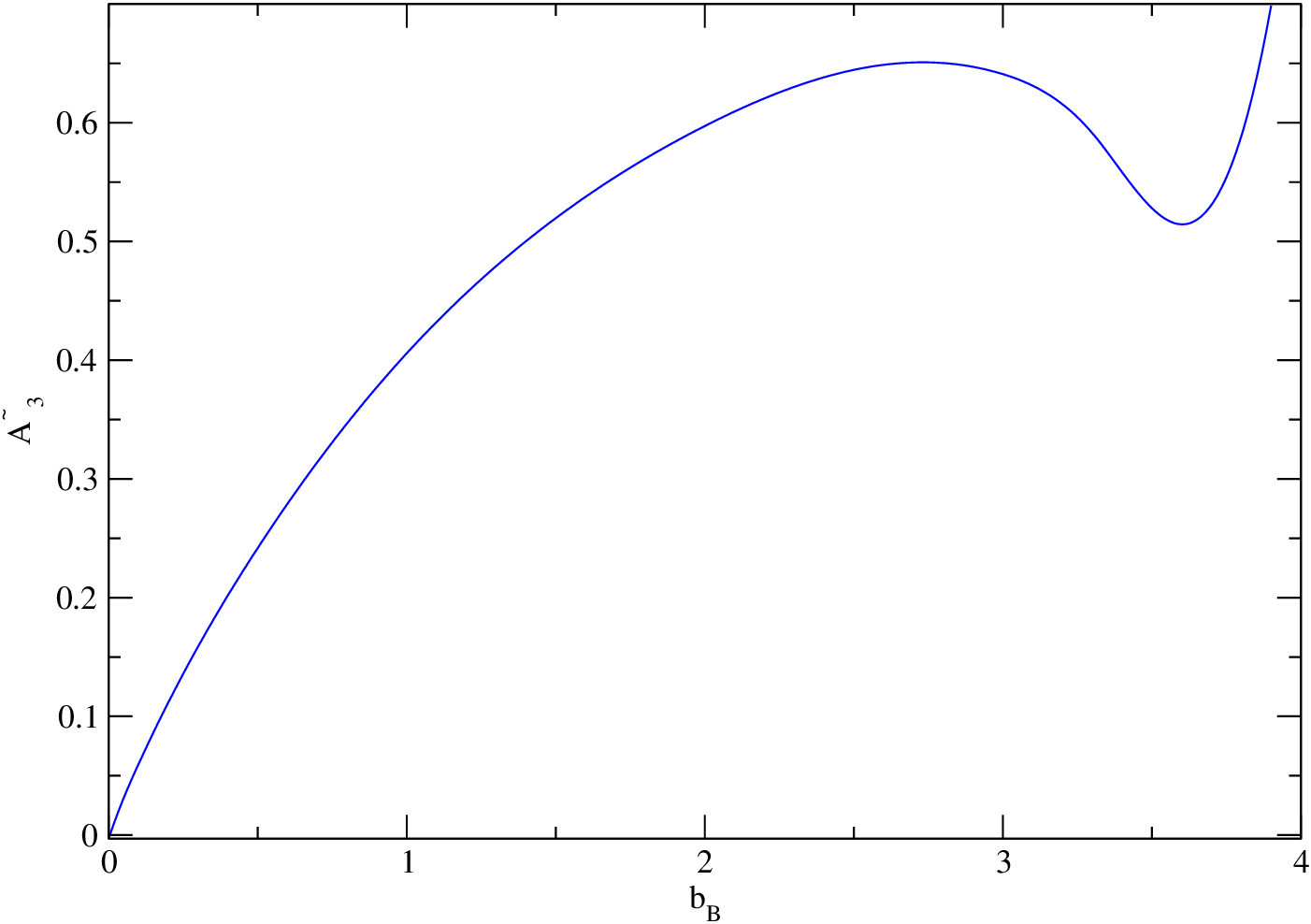}
\hskip 1.0cm 
\epsfxsize=7.0cm\epsfysize=6.0cm
                      \epsfbox{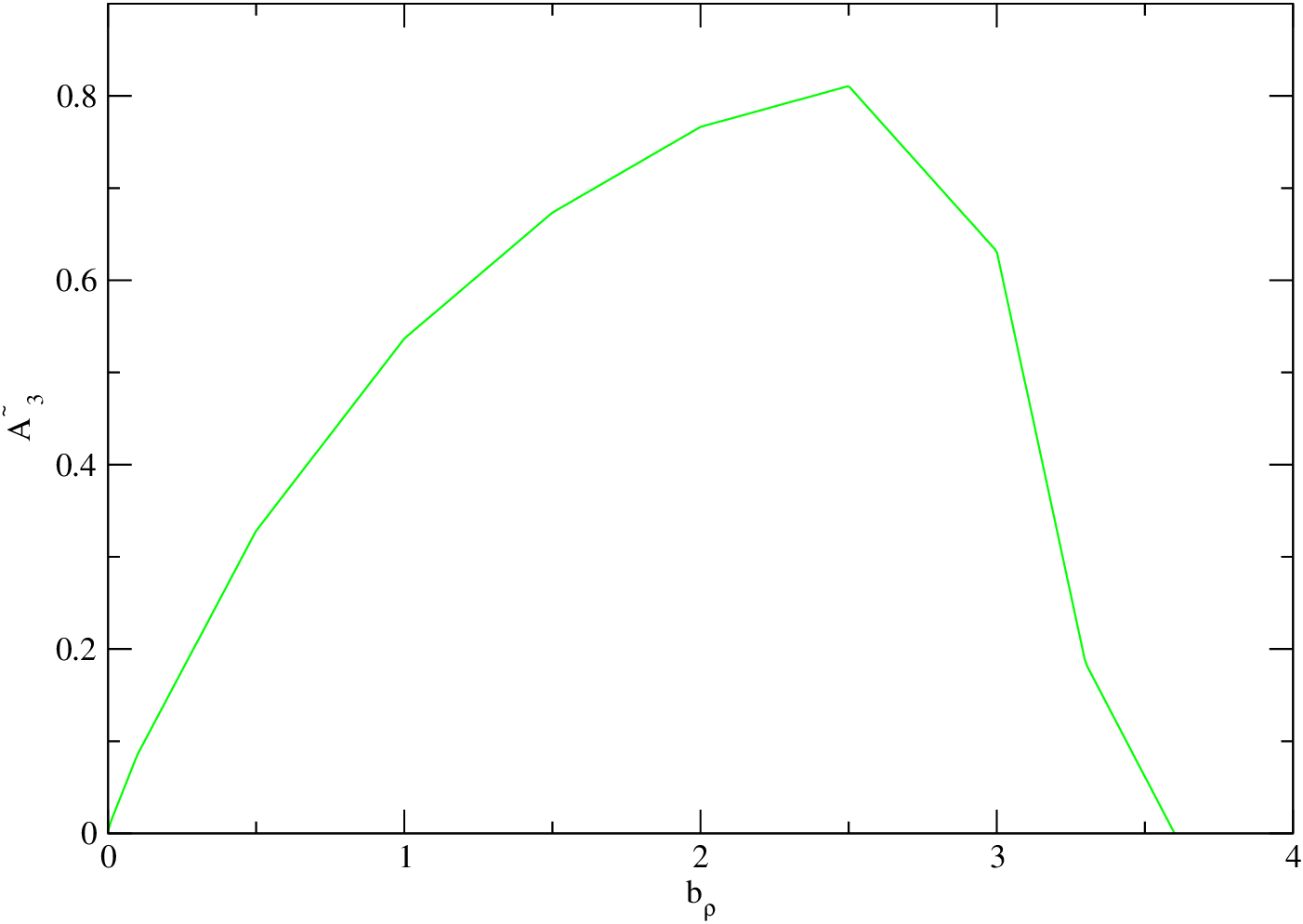}}
\caption{Variation of ${\tilde{A}}_3$ with $\bb$ (left) and $\brho$ (right).}
\end{figure}
%

\begin{thebibliography}{}
\bibitem{ali} A.~Ali, hep-ph/0312303; R.~Fleischer, hep-ph/0405091;
  H-n.~Li, hep-ph/0303116.

\bibitem{neubert} M.~Neubert, Phys. Rept.{\bf 245}, 260 (1994).

\bibitem{fac} M.~Bauer, B.~Stech and M.~Wirbel, Z. Phys. {\bf C34}, 103
(1987); M.~Bauer, B.~Stech and M.~Wirbel, Z. Phys {\bf C29}, 637 (1985).

\bibitem{li1} H.~Y.~Cheng, H-n.~Li and K.~C.~Yang, Phys. Rev. {\bf D60},
094005 (1999).

\bibitem{li2} T.~W.~Yeh and H-n.~Li, Phys. Rev. {\bf D56}, 1615 (1997).

\bibitem{bbns}M.~Beneke, G.~Buchalla, M.~Neubert and C.~T.~Sachrajda,
Phys. Rev. Lett. {\bf 83}, 1914 (1999); M.~Beneke, G.~Buchalla, M.~Neubert
and C.~T.~Sachrajda, Nucl. Phys. {\bf B591}, 313 (2000).

\bibitem{pqcd} G.~P.~Leepage and S.~J.~Brodsky, Phys. Rev. {\bf 22}, 2157
(1980); A.~V.~Efremov and A.~V.~Radyushkin, Theor. Math. Phys. {\bf 42}, 97
(1980); A.~V.~Efremov and A.~V.~Radyushkin, Phys. Lett. {\bf B94}, 245
(1980); A.~Duncan and A.~H.~Mueller, Phys. Lett. {\bf B90}, 159 (1980).

\bibitem{brodsky} A.~Szczepaniak, E.~M.~Henley and S.~J.~Brodsky,
Phys. Lett. {\bf B243}, 287 (1990). 

\bibitem{hwang}D.~S.~Hwang and B-H.~Lee, Eur. Phys. J. {\bf C6}, 663 (1999).

\bibitem{sanda}T.~Kurimoto, H-n.~Li and A.~I.~Sanda, Phys. Rev. {\bf D65},
014007 (2001).

\bibitem{collins}J.~C.~Collins, D.~E.~Soper and G.~Sterman in {\it
Perturbative Quantum Chromodynamics}, ed. A.~H.~Mueller, World
Scientific (1989).

\bibitem{isgur} N.~Isgur and C.~H.~Llewellyn Smith,
Phys. Rev. Lett. {\bf 52}, 1080 (1984); N.~Isgur and C.~H.~Llewellyn Smith, 
Phys. Lett. {\bf B217}, 535 (1989); N.~Isgur and C.~H.~Llewellyn Smith,
Nucl. Phys. {\bf B381}, 129 (1992); A.~V.~~Radyushkin, Nucl. Phys. 
{\bf A532}, 141 (1991).

\bibitem{sterman}H-n.~Li and G.~Sterman, Nucl. Phys. {\bf B381}, 129 (1992).

\bibitem{ball1}P.~Ball and V.~M.~Braun, Phys. Rev. {\bf D55}, 5561 (1997).

\bibitem{botts}J.~Botts and G.~Sterman, Nucl. Phys. {\bf B325}, 62 (1989).

\bibitem{li3}H-n.~Li, Phys. Rev. {\bf D66}, 094010 (2000).

\bibitem{sachrajda}S.~Descotes-Genon and C.~T.~Sachrajda,
Nucl. Phys. {\bf B625}, 239 (2002).

\bibitem{lepage}S.~J.~Brodsky and G.~P.~Lepage in {\it
Perturbative Quantum Chromodynamics}, ed. A.~H.~Mueller, World
Scientific (1989).  

\bibitem{ball2}P.~Ball and V.~M.~Braun, hep-ph/9808229.

\bibitem{beneke}M.~Beneke and T.~Feldmann, Nucl. Phys. {\bf B592}, 3 (2001).

\bibitem{wei}Z.-T.~Wei and M.-Z.~Yang, Nucl. Phys. {\bf B624}, 263 (2002). 

\bibitem{li4}H-n.~Li and H.~Yu, Phys. Rev. {\bf D53}, 2480 (1996).

\bibitem{li5}H-n.~Li, Phys. Rev. {\bf D52}, 3958 (1994).

\bibitem{frozen}A.~C.~Aguilar, A.~Mihara and A.~A.~Natale,
 Phys. Rev. {\bf D65}, 054011 (2002).

\bibitem{cornwall}J.~M.~Cornwall, Phys. Rev. {\bf D26}, 1453 (1982).

\bibitem{ball3}P.~Ball and V.~M.~Braun, Phys. Rev. {\bf D58}, 094016 (1998). 
\end{thebibliography}
\end{document}